\documentclass{svjour3}

\smartqed  
\usepackage{amsmath,amssymb}
\usepackage{graphicx}
\usepackage{natbib}
\bibpunct{(}{)}{;}{a}{}{,}

\def\ni{\noindent}
\def\be{\begin{equation}}
\def\ee{\end{equation}}

\newcommand{\beq}{\begin{eqnarray}}
\newcommand{\eeq}{\end{eqnarray}}

\newcommand{\aeta}[3]{  #1, {A\&A}, {  #2}, #3}

\newcommand{\araa}[3]{ #1, {ARAA}, { #2}, #3}
\newcommand{\apj}[3]{  #1, {ApJ}, { #2}, #3}
\newcommand{\pasj}[3]{  #1, {PASJ}, 
{ #2}, #3}

\journalname{SSRv}

\begin{document}

\title{Nonthermal phenomena in clusters of galaxies}

\author{Y.~Rephaeli \and
      J.~Nevalainen \and
      T.~Ohashi \and
      A.M.~Bykov
}

\authorrunning{Rephaeli et al.}

\institute{Y. Rephaeli \at
            School of Physics \& Astronomy, Tel Aviv University, 
                    Tel Aviv, 69978, Israel\\
            Center for Astrophysics and Space Sciences, University of 
            California, San Diego, La Jolla, CA\,92093-0424\\ 
            \email{yoelr@wise.tau.ac.il}
         \and
         J. Nevalainen \at
            Observatory, P.O. Box 14, 00014 University of Helsinki, Helsinki, Finland\\
         \and
         T. Ohashi \at 
            Department of Physics, Tokyo Metropolitan University, 
            1-1 Minami-Osawa, Tokyo 192-0397, Japan\\
         \and
         A.M. Bykov \at
            A.F. Ioffe Institute for Physics and Technology, 194021
            St. Petersburg, Russia} 

\date{Received: 1 October 2007; Accepted: 5 December 2007 }

\maketitle

\begin{abstract}
Recent observations of high energy ($> 20$ keV) X-ray emission in a few 
clusters of galaxies broaden our knowledge of physical phenomena in the intracluster 
space. This emission is likely to be nonthermal, probably resulting from 
Compton scattering of relativistic electrons by the cosmic microwave 
background (CMB) radiation. Direct evidence for the presence of relativistic 
electrons in some $50$ clusters comes from measurements of extended 
radio emission in their central regions. We briefly review the main results 
from observations of extended regions of radio emission, and Faraday rotation 
measurements of background and cluster radio sources. The main focus of the 
review are searches for nonthermal X-ray emission conducted with past and 
currently operating satellites, which yielded appreciable evidence for 
nonthermal emission components in the spectra of a few clusters. This 
evidence is clearly not unequivocal, due to substantial observational and 
systematic uncertainties, in addition to virtually complete lack of spatial 
information. If indeed the emission has its origin in Compton scattering of 
relativistic electrons by the CMB, then the mean magnetic field strength and 
density of relativistic electrons in the cluster can be directly determined. Knowledge 
of these basic nonthermal quantities is valuable for the detailed description 
of processes in intracluster gas and for the origin of magnetic fields.

\keywords{Clusters: general, X-ray emission}

\end{abstract}

\section{Introduction}

Clusters of galaxies are the largest bound systems and the most important 
link to the large scale structure (LSS) of the Universe. The detailed 
properties of clusters, such as the distributions of their various mass 
constituents - dark matter, hot intracluster (IC) gas, and galaxies - 
dynamics, and thermal structure, are of much interest both intrinsically, 
and for the understanding of the formation and evolution of the LSS. 
Moreover, detailed {\sl astrophysical} knowledge of clusters is essential 
for their use as {\sl precise} cosmological probes to measure global 
parameters, such as $H_0$, $\Omega_{\mathrm M}$, \& $\Omega_{\Lambda}$, and 
parameters characterising the primordial density fluctuation field.

Recent observations of many clusters of galaxies with the {\sl Chandra} 
and {\sl XMM} satellites at energies $\epsilon \leq 10$ keV, have 
significantly advanced our knowledge of the morphology and thermal 
structure of hot IC gas, the source of the cluster thermal Bremsstrahlung 
emission. The improved determinations of the gas temperature, density, 
and metal abundances from these observations significantly improve 
estimates of such important quantities as the total cluster mass and its 
gaseous and baryonic mass fractions. 

As has been the case in galaxies, in clusters too a more physically 
complete understanding of these systems necessitates knowledge also of 
non-thermal (NT) quantities and phenomena in the IC space. Observational 
evidence for the relevance of these phenomena in clusters comes mostly 
from measurements of extended regions of radio emission, and from Faraday 
rotation (FR) of the plane of polarisation of radio sources seen through 
(or inside) clusters. Since the observed radio emission is clearly 
synchrotron-produced, its level and spectrum yield direct information on 
IC relativistic electrons and magnetic fields. Information on cluster magnetic 
fields (separately from relativistic electron properties) is obtained also from 
FR measurements. Compton scattering of cosmic microwave background (CMB) 
photons by the radio-emitting relativistic electrons boosts photon energies 
to the X-and-$\gamma$ regions (e.g., \citealt{rephaeli1979}). The search for 
cluster NT X-ray emission has begun long ago \citep{rephaeligruber1987}, but 
first clear indications for emission at energies $\epsilon \geq 20$ keV 
came only after deep dedicated observations of a few clusters with the 
{\sl{RXTE}} and {\sl{BeppoSAX}} satellites (beginning with analyses of observations of the 
Coma cluster \citep{rephaeli1999,fuscofemiano1999}. Radio and NT 
X-ray observations provide quantitative measures of very appreciable 
magnetic fields and relativistic electron densities in the observed clusters. 
These results open a new dimension in the study of clusters. 

This is a review of cluster NT X-ray observations and their direct 
implications, including prospects for the detection of $\gamma$-ray 
emission. The literature on cluster NT phenomena is (perhaps somewhat 
surprisingly) quite extensive, including several reviews of radio 
emission (e.g., \citealt{govoni2004}) and cluster magnetic fields 
(e.g., \citealt{carilli2002}), and a review of the current status of 
radio observations by \citealt{ferrari2008} - Chapter 6, this volume. In 
order to properly address the comparison between magnetic field values 
deduced from radio observations and jointly from NT X-ray and radio 
measurements, we include here a brief summary of cluster radio observations. 
Measurements of EUV emission in a few clusters, and claims that this 
emission is by energetic electrons, are reviewed by \citealt{durret2008} - 
Chapter 4, this volume. NT radiation processes are reviewed by \citealt{petrosian2008a} - Chapter 10, this volume, and relevant aspects 
of particle acceleration mechanisms are reviewed by \citealt{petrosian2008b} - Chapter 11, this volume.

\section{Radio Observations}

Measurements of synchrotron radio emission at several frequencies usually 
provide the first evidence for the presence of a significant population of 
relativistic electrons and magnetic fields. This has been the prime evidence for 
IC fields and electrons. In addition, Faraday Rotation (FR) of the plane 
of polarisation of radiation from cluster and background radio galaxies 
has provided mostly statistical evidence for IC magnetic fields. Main 
results from these very different sets of observations are briefly 
reviewed in the next two subsections.

\subsection{Extended IC Emission}

In clusters the task of actually determining that the observed emission 
is from extended IC regions is quite challenging, since high resolution 
observations of the various discrete radio sources (that are mostly in 
the member galaxies) are required, in addition to the (lower resolution) 
observations of the extended low brightness emission. The truly extended 
emission is mapped upon subtraction of the respective contributions of 
these sources to the spectral flux. Observations of extended radio 
emission had begun some 50 years ago with measurements of Coma C in the 
central region of the Coma cluster by \citet{seeger1957}, followed by 
various other measurements at frequencies in the range $0.01-4.85$ GHz, 
including the first detailed study at $408$ MHz and $1407$ MHz by 
\citet{willson1970}.

Appreciable effort has been devoted to measure extended emission regions 
in clusters, which has resulted in the mapping of centrally located 
(`halo') and other (`relic') regions in about 50 clusters. Some 32 of 
these were found in a VLA survey \citep{giovannini1999,giovannini2000} of 205 nearby 
clusters in the ACO catalogue, only about a dozen of these were previously 
known to have regions of extended radio emission. Primary interest 
(certainly to us here) is in the former sources, which constitute a truly 
cluster phenomenon. A central extended radio region (which is somewhat 
inappropriately referred to as `halo') typically has a size of $\sim 1-2$ 
Mpc, and a luminosity in the range $10^{40} - 10^{42}$ erg\,s$^{-1}$ 
(for $H_0 = 70 \;{\rm km}\;{\rm s}^{-1} \;{\rm Mpc}^{-1}$) over the frequency band 
$\sim 0.04-5$ GHz. With radio indices usually in the range 
$\sim 1-2$, the emission is appreciably steeper than that of (most) 
radio galaxies. 

The clusters in which radio halos and relics have already been found 
include evolving systems with a substantial degree of subclustering, as 
well as well-relaxed clusters that seem to have attained hydrostatic 
equilibrium. Halo morphologies are also quite varied, from roughly 
circular (projected) configuration to highly irregular shape, as can 
be seen in the contour maps produced by \citet{giovannini1999,giovannini2000,giovannini2006}. The spatial variety is reflected also in the wide distribution 
of values of the spectral index across the halo; e.g. see the maps in 
\citet{giovannini1999}.

Obviously, the magnetic field strength and relativistic electron density 
cannot both be determined from radio measurements alone, unless it 
is assumed that they are related so both can be determined from a 
single observable. It is commonly assumed that the total energy density 
of particles (mostly, protons and electrons) is equal to that in the 
magnetic field. The validity of this equipartition assumption is not 
obvious, especially in clusters where the particles and fields may have 
different origins and evolutionary histories. Moreover, the complex 
nature of the radio emission, and the expected significant variation 
of the magnetic field strength and relativistic electron density across a halo, 
imply that only rough estimates of these quantities can be obtained 
from measurements of the (spectral) flux integrated over the halo. 
When available, values of the halo mean equipartition field, $B_{\mathrm{eq}}$, 
are also listed in Table 1; generally, these substantially uncertain 
values are at the few $\mu$G level.

\subsection{Faraday Rotation}

IC magnetic fields can also be estimated by measuring the statistical 
depolarisation and Faraday Rotation of the plane of polarisation of 
radiation from background radio sources seen through clusters (e.g. \citealt{kim1991}), and also from radio sources in the cluster. FR measurements sample 
the line of sight component of the randomly oriented (and spatially 
dependent) IC fields, weighted by the gas density, yielding a mean 
weighted value which we denote by $B_{\mathrm{fr}}$. This quantity was estimated 
by analysing the rotation measure (RM) distribution of individual radio 
sources in several clusters, including Cygnus~A \citep{dreher1987}, 
Hydra~A \citep{taylor1993,vogt2003}, A~119 \citep{feretti1999,dolag2001}, A~400, and A~2634 \citep{vogt2003}. Analyses 
of FR measurements typically yield values of $B_{\mathrm{fr}}$ that are in the 
range of $1-10$ $\mu$G.

Most FR studies are statistical, based on measurements of radio sources 
that are inside or behind clusters. An example is the work of \citet{clarke2001}, who determined the distribution of RM values with cluster-centric 
distance for 27 radio sources within or in the background (15 and 12, 
respectively) of 16 nearby clusters. Analysis of this distribution 
(including comparison with results for a control sample of radio sources 
seen outside the central regions of the clusters in the sample) yielded 
a mean field value of $\sim 5-10\, (\ell/10\ {\rm kpc})^{-1/2}$ 
$\mu$G, where $\ell$ is a characteristic field spatial coherence 
(reversal) scale. In further work the sample of radio sources was 
significantly expanded (to about 70; \citealt{clarke2004}). In comparing 
different measures of the mean strength of IC fields it should be 
remembered that the selective sampling of locally enhanced fields in 
high gas density regions in cluster cores broadens the RM distribution, 
resulting in overestimation of the field mean strength.

Deduced values of $B_{\mathrm{fr}}$ yield substantially uncertain estimates of 
the mean field across a halo. The major inherent uncertainties stem from 
the need to separate the several contributions to the total RM (including 
that which is intrinsic to the radio source), the unknown tangled 
morphology of IC fields and their spatial variation across the cluster, 
as well as uncertainty in modelling the gas density profile. A discussion 
of these uncertainties, many aspects of which have already been assessed 
in some detail (e.g., \citealt{goldshmidt1993,newman2002,rudnick2003,ensslin2003,murgia2004}), is out 
of the scope of our review. 

\section{Nonthermal X-ray Emission}

As has been noted long ago, Compton scattering of the CMB by relativistic 
electrons boosts photon energies to the X-ray and $\gamma$-ray 
regions. Measurement of this radiation provides additional information 
that enables the determination of the electron density and mean magnetic 
field directly, {\sl without} the need to invoke equipartition. The mean 
strength of the magnetic field which is deduced from the radio and NT 
X-ray fluxes, $B_{\mathrm{rx}}$, is essentially a volume average over the emitting 
region. The detection of IC radio and NT X-ray emission sets the stage 
for a more meaningful study of the origin of magnetic fields and 
cosmic rays in extragalactic environments. 

The focus of this review are measurements of NT X-ray emission in 
large extended IC regions. NT emission is obviously predicted within 
dominant radio galaxies in the centres of clusters and in their radio 
halos. This emission has been sought in several galaxies including 
NGC~1275 in the Perseus cluster (e.g., \citealt{sanders2007}) and M~87 
in Virgo \citep{simionescu2007}, but does not seem to have been detected. 
The mapping of the electron spectrum in these inner regions clearly 
provides crucial information on the initial source of IC relativistic electrons 
and most likely also energetic protons, whose energy input and heating 
of the gas could possibly be important in the inner core of Perseus 
\citep{rephaeli1995} and other clusters. NT X-ray emission in these 
environments is quite complex due to its multi-source (AGN, jet, 
binaries, and halo), temporally variable nature. A discussion of 
NT emission in these essentially galactic environments is out of the 
scope of our review. 

The prospects for measuring cluster NT X-ray emission motivated detailed 
calculations of the predicted emission \citep{rephaeli1977,rephaeli1979}. In these 
calculations the relativistic electron spectrum was directly related to the 
measured radio spectrum and no attempt was made to model the spectra of 
possible energetic electron populations. Energetic protons, which are a 
major Galactic cosmic ray component, are also expected in the IC space, 
particularly so in the inner cores of clusters dominated by a radio 
galaxy. Their interactions with protons in the gas produce neutral and 
charged pions, whose decays yield $\gamma$-ray emission and secondary 
electrons (e.g., \citealt{dennison1980,dermer1988,blasi1999}). Energetic protons also deposit energy and heat IC 
gas through their Coulomb interactions with electrons and protons in the 
gas \citep{rephaeli1987,rephaeli1995}. 

With mean field values in the range of $0.1 -1$ $\mu$G, the energy 
range of electrons emitting at the observed radio frequencies is 
$\sim 1-100$ GeV. Of course, electrons with energies outside this 
range are also expected, either as part of the same or a different 
population. In particular, NT Bremsstrahlung EUV and X-ray emission by 
(the more numerous) lower energy electrons would also be expected 
\citep{kaastra1998,sarazin1999}. However, at energies below 
$\sim 200$ MeV, the main energy loss is electronic excitations in 
the gas \citep{rephaeli1979}; this sets a stringent limit on the contribution 
to the NT X-ray emission by a low energy electron population \citep{rephaeli2001,petrosian2001}. We will therefore assume that NT X-ray emission 
is largely due to Compton scattering of radio-emitting electrons by 
the CMB. 

The main characteristics of the predicted Compton-produced emission 
are: (a) X-ray to radio flux ratio is roughly equal to the ratio 
of the CMB energy density to the magnetic field energy density. (b) 
Power-law index is nearly equal to the radio index. (c) Matching 
X-ray and radio centroids, with the X-ray spatial profile generally 
shallower than that of the radio emission. These features are expected 
only if the Compton-produced emission is identified and separated from 
other contributions, the primary thermal emission as well as NT emission 
from relics and individual galaxies. Due to source confusion, as well as 
other systematic and observational uncertainties, spatial information is 
{\sl crucially} needed in order to clearly identify the origin of NT 
X-ray emission. 

We first briefly summarise the required sensitivity and level of spatial 
resolution for the detection of NT emission, and then review the results 
of the search for this emission with the {\sl HEAO-1}, {\sl CGRO}, 
{\sl RXTE}, {\sl BeppoSAX}, {\sl{Suzaku}}, and {\sl INTEGRAL}.

\subsection{Instrumental Requirements}

Cluster X-ray emission is primarily thermal up to $\sim 30-40$ keV; 
therefore, clear identification of NT emission necessitates also precise 
measurement of the thermal emission in order to account for and separate 
it from the sought NT component. The required detector spectral range 
must therefore extend down to sufficiently low energies.

Using standard expressions for Compton and synchrotron emission from 
the same population of relativistic electrons, the predicted level of 
Compton-produced emission from nearby clusters with measured radio halos 
can be readily estimated. Since imaging information at the high energy 
X-ray regime ($\epsilon > 30$) is minimal at best, we will ignore spatial 
factors in the Compton-synchrotron formulae (e.g., \citealt{rephaeli1979}) that 
include integrals over the relativistic electron and magnetic fields spatial 
profiles. This is valid as long as the NT X-ray emission region 
coincides with the radio region. We base our estimate of the required 
detector sensitivity on the level of radio emission as measured in the 
nearby clusters with well measured halos (such as Coma, A~2256, A~2319), for 
which the feasibility of detection of the NT emission is likely to 
be highest. Using the range of values for radio spectral index and 
flux, we estimate the level of NT flux at 40 keV to be typically 
$\sim 1\times 10^{-6}$ cm$^{-2}$\, s$^{-1}$\,keV$^{-1}$ and an 
integrated flux of $\sim 3\times 10^{-5}$ cm$^{-2}$\, s$^{-1}$ in the 
$40-80$ keV band, if the mean field (across the emitting region) is 
$B \sim 0.3$ $\mu$G. The value of the field is critical due to the 
strong $B$ dependence of the flux, with a power-law exponent typically 
larger than 2. For example, the predicted flux in the Coma cluster is 
some $16$ times lower if $B = 1$ $\mu$G. 

Given the great difficulty in detecting the predicted NT flux at 
energies $\epsilon \geq 40$ keV, it is only reasonable to search for 
it also at lower energies. To do so optimally would require a 
high-sensitivity detector that covers a wider spectral band than 
that of any previously flown instrument. Combining measurements made 
with two different detectors is problematic, given the inherent 
difficulty in the requisite precise cross-calibration (at the 
$\sim 1$~\% level) of signals from detectors on the same satellite, 
and even more so when the detectors were on different satellites 
altogether. An added difficulty arises in the extremely difficult 
task of detecting a weak secondary component at energies below 
$\sim 1$ keV where photoelectric absorption is strong. This should 
be kept in mind when assessing results form such combined analyses, 
some of which are discussed below.

Even if NT emission is detected its origin has to be verified before 
it is identified with electrons in the halo. The first requirement is 
that the measured emission does not show any temporal variation (that 
would usually imply AGN origin). Sizes of radio halos in nearby clusters 
are in the range of $15' - 30'$, which in principle can be resolved 
by the {\sl IBIS} instrument aboard {\sl INTEGRAL}, the first imaging 
experiment in the high energy X-ray regime. Clearly, higher detector 
sensitivity is required for resolved measurements of such large regions. 
All the other presently active (and previous) satellites with high energy 
spectral capability have FOVs that are larger than the halos of (even) 
nearby clusters. Measurements of NT emission have so far been solely 
spectral, thus providing only the necessary - but not sufficient - 
evidence for the detection of NT emission in a few clusters. 

\subsection{Initial search for NT X-ray emission}

Systematic searches for NT X-ray emission began with the analysis of 
archival {\sl HEAO-1} measurements of six clusters (Coma, A~401, A~2142, 
A~2255, A~2256, A~2319) with measured radio halos \citep{rephaeligruber1987,rephaeli1988}, and continued with {\sl CGRO} 
observation of the Coma cluster \citep{rephaeli1994}. No significant NT 
emission was detected, resulting in lower limits on the mean, 
volume-averaged magnetic fields in these clusters, $B_{\mathrm{rx}}= O(0.1\; \mu$G).

Observation of the Coma cluster with the {\sl CGRO/OSSE} experiment for 
$\sim 380$ ks was the first dedicated measurement of a cluster aimed at 
measuring NT emission. The fact that the predicted emission was not 
detected clearly established that instruments with higher detector 
sensitivity, significantly lower level of internal background, and 
much smaller FOV (than that of {\sl OSSE}, $\sim 1^{\circ} 
\times 1^{\circ}$ FWHM), are the minimal requirements for 
measurement of emission in the $40-100$ keV band. 

Attempts to measure NT emission from clusters continued with all 
subsequent X-ray satellites whose nominal spectral band extended beyond 
10 keV. First observational evidence for NT emission came from long 
measurements with the {\sl RXTE} and {\sl BeppoSAX} satellites, as 
discussed in the next two subsections, followed by a review of results 
from {\sl{ASCA}}, initial results from {\sl Suzaku/HXD}, and the first 
spatial analysis of {\sl INTEGRAL/IBIS} observations of the Coma 
cluster. 

\subsection{Search for NT emission with {\sl RXTE}}

The search for cluster NT emission advanced significantly through long 
observations with the PCA and HEXTE experiments aboard the {\sl RXTE} 
satellite. These instruments have the minimally required capabilities 
for detection of a weak NT spectral component - sufficiently 
high sensitivity for precise measurement of the primary thermal emission, 
and good background subtraction achieved by (on and off source) 
`rocking' of the HEXTE detectors. The crucial {\sl RXTE} features which 
are essential for identifying the small NT component are spectral 
overlap of the two experiments over the effective (narrower than 
nominal) $\sim 13-25$ keV band, and the same triangular spatial response 
function with $58'$ FWHM. As has been noted already, the {\sl RXTE} 
does not have the capability to resolve the emission even in nearby 
(rich) clusters, implying that the exact origin of NT emission cannot be 
identified even when a secondary spectral component is clearly detected.

We review here results of searches for NT components in the spectra of 
clusters that were observed for at least $\sim 100$ ks. Shorter  
observations were mostly aimed at measurement of the primary thermal 
emission, with the exception of a $70$ ks observation of A~754 \citep{valinia1999}, and even shorter $\sim 30$ ks observations of A~2256 \citep{henriksen1999} and A~1367 \citep{henriksen2001}, which resulted in upper 
limits on NT emission in these clusters. For obvious reasons, Coma was 
the first cluster searched for NT emission with {\sl RXTE} \citep{rephaeli1999}. The search continued with long observations of A~2319, A~2256, 
and A~2163, all with observed radio halos. Results of the search for NT 
emission in these and other clusters are listed in Table \ref{hxr_xte.tab}, 
and briefly described below.

\begin{table}
\caption{NT emission parameters from {\sl{RXTE}} measurements and (deduced 
mean radio-and-X-determined field) $B_{\mathrm{rx}}$}.
\label{hxr_xte.tab}
\vspace{0.2cm}
\begin{tabular}{lllll}
\hline
Cluster & $20-80$ keV flux          & $\Gamma$         & $B_{\mathrm{rx}}$    & Reference\\
        & ($10^{-12}$ erg\,s$^{-1}$\,cm$^{-2}$) &     & ($\mu$G)    &\\
\hline
Coma          & $21 \pm 6$       &  $2.1\pm 0.5$       & $0.1-0.3$           & \protect\citet{rephaeli2002}\\
A~2163        & $11_{-9}^{+17}$  & $1.8_{-4.2}^{+0.9}$ & $0.4\pm 0.2$        & \protect\citet{rephaeli2006}\\
A~2256        & $4.6\pm 2.4$     & $2.2_{-0.3}^{+0.9}$ & $0.2^{+1.0}_{-0.1}$ & \protect\citet{rephaeli2003}\\
A~2319        & $14\pm 3$        & $2.4\pm 0.3$        & $0.1-0.3$           & \protect\citet{gruber2002}\\
A~3667        & $\leq 4$         & $\sim 2.1$          & $\geq 0.4$          & \protect\citet{rephaeli2004}\\
1ES 0657-55.8 & $\sim 5\pm 3$    & $1.6\pm 0.3$        & $1.2^{\bf a}$       & \protect\citet{petrosian2006}\\
\hline
\multicolumn{5}{l}{All quoted errors are at the 90~\% confidence level.}\\
\multicolumn{5}{l}{$^{\bf a}$ The magnetic field value in 1ES 0657-558
was derived assuming energy}\\
\multicolumn{5}{l}{equipartition between the field and particles.}\\
\end{tabular}
\end{table} 

\ni{\bf{Coma:}} The cluster was initially observed (in 1996) for 
$\sim 90$ ks with the PCA, and for $\sim 29$ ks with HEXTE. 
Analysis of these measurements showed evidence for the presence 
of a second spectral component at energies up to $\sim 20$ keV 
\citep{rephaeli1999}, but the nature of the secondary emission could not 
be determined. In order to improve the quality of the spectral results, 
the cluster was observed again (in 2000) for $\sim$177\,ks. Analysis 
of the new observations (\citealt{rephaeli2002}, hereafter RG02) yielded 
results that are consistent with those from the first observation. Joint 
analysis of the full {\sl RXTE} dataset shows that - from a statistical 
point of view - no preference could be determined for the nature of the 
secondary component. If thermal, the most probable temperature would be 
very high, ${\mathrm k}T_2 \simeq 37.1$ keV, but at its lowest boundary, the 90~\% 
contour region includes a much lower, physically more acceptable value. 
However, the best fit combination of the temperatures of the primary and 
secondary components is then ${\mathrm k}T_{1}=5.5$ keV, and ${\mathrm k}T_{2}=9$ keV, 
respectively, with the respective $4-20$ keV fractional fluxes of $\sim 
24~\%$ and $\sim 76~\%$. RG02 argued that it is quite unlikely that about 
a quarter of the emission comes from gas at a significantly lower temperature 
than the mean value deduced by virtually all previous X-ray satellites. In 
particular, such a component would have been detected in the high spatially 
resolved {\sl XMM} and {\sl Chandra} maps of the cluster. 

\begin{figure}
\begin{center}
\includegraphics[width=7cm,angle=270]{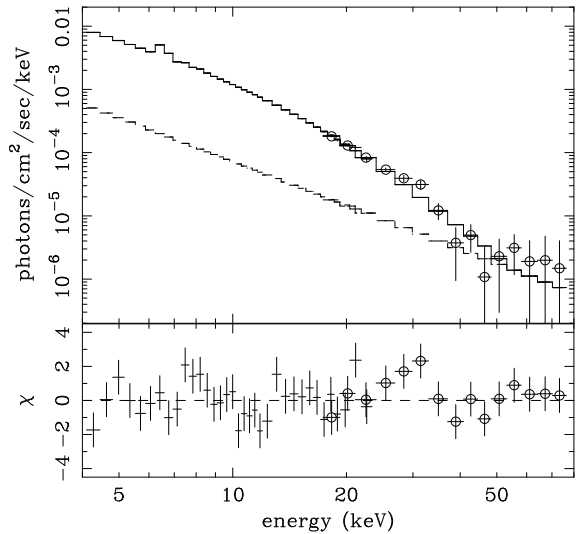}
\caption{The RXTE spectrum of the Coma cluster and folded Raymond-Smith 
(${\mathrm k}T \simeq 7.67$), and power-law (photon index $=2.3$) models (from 
\protect\citealt{rephaeli2002}). HEXTE data points are marked with circles and 
68~\% error bars. The total fitted spectrum is shown with a histogram, 
while the lower histogram shows the power-law portion of the best fit. 
The quality of the fit is demonstrated in the lower panel, which displays 
the observed difference normalised to the standard error of the data point.}
\label{fig:fig1}
\end{center}
\end{figure}    

To quantitatively assess the possibility that a two-temperature gas model is
just a simplified representation of a more realistic continuous temperature
distribution, RG02 have assumed a polytropic gas temperature profile of the form
$T(r) \propto n(r)^{\gamma -1}$, with a $\beta$ profile for the gas density,
$n(r) \propto (1+r^{2}/r_{\mathrm c}^2)^{-3\beta/2}$, where $r_{\mathrm c}$ is
the core radius. With these profiles they calculated the integrated flux and the
mean emissivity-weighted temperatures as functions of $\gamma$, $\beta$, and
$r$. These quantities were then calculated in the projected (2D) circular
regions $[0, R]$ and $[R, R_0]$ by convolving over the triangular response of
the PCA with $R_0 \simeq 58'$. From ROSAT observations,  $r_{\mathrm c} \sim
10.0'$, and $\beta \simeq 0.70 \pm 0.05$ \citep{mohr1999}. The range of values
of $R$, $\beta$, and $\gamma$ for which the two mean emissivity-weighted
temperatures and respective fluxes from these regions were closest to the values
deduced from the spectral analysis, were then determined. The results of these
calculations indicate that for $0.5 \leq \beta \leq 0.9$  and $1 \leq \gamma
\leq 5/3$, there is no acceptable polytropic configuration that matches the
observationally deduced values of the temperatures and fractional fluxes. For
low values of $\gamma$ the temperature gradient is too shallow, whereas for high
$\gamma$ values the implied central temperature is unrealistically high. Based
on these results RG02 concluded that the two thermal components model does not
seem to be consistent with the {\sl{RXTE}} measurements, but certainly could not
be ruled out.

Clearly, a search for NT emission motivated the long observation, and 
this alternative origin was assessed in detail by RG02. Since emission 
from an AGN in the FOV was considered unlikely, they assessed whether the 
secondary emission is due to Compton scattering of relativistic electrons 
whose presence in Coma is directly inferred from many measurements of 
spatially extended region of radio emission (e.g. \citealt{kim1991,giovannini1993,thierbach2003}). From the measured radio spectral 
index, $1.34\pm 0.1$, it follows that the predicted power-law (photon) 
flux from Compton scattering of these electrons by the CMB should have an 
index $\Gamma \sim 2.3$. A power-law fit to the secondary 
component yielded $2.1 \pm 0.5$ at the 90~\% confidence level (CL). 
Results of this fit are shown in Fig.~\ref{fig:fig1}.

With the measured mean radio flux of $0.72 \pm 0.21$ Jy at 1 GHz, 
and the deduced level of power-law X-ray flux, the mean volume-averaged 
value of the magnetic field, $B_{\mathrm{rx}}$, was computed to be in the range 
$0.1 - 0.3$~$\mu$G. As noted by RG02, this estimate is based on the 
assumption that the spatial factors in the theoretical expressions for 
the two fluxes \citep{rephaeli1979} are roughly equal, an implicit assumption 
made in virtually all previous attempts to derive the strength of cluster 
fields from radio and X-ray measurements. 

A similar procedure was employed in analyses of observations of the 
other clusters (A~2319, A~2256, and A~2163) that were initiated by Rephaeli 
\& Gruber.

\ni{\bf{A~2319:}} The cluster was observed (in 1999) for $\sim 160$ ks. 
Analysis of the data \citep{gruber2002} showed no  noticeable 
variability over the $\sim 8$ week observation. The quality of the 
data allowed a meaningful search for emission whose spectral 
properties are distinct from those of the primary thermal emission 
with measured mean temperature in the range $8-10$ keV. 
A fit to a single thermal component yielded ${\mathrm k}T = 8.6 \pm 0.1$ 
(90~\% CL), a low iron abundance $Z_{\mathrm{Fe}} \sim 0.16 \pm 0.02$, 
and large positive residuals below 6 keV and between 15 to 30 keV. 
The quality of the fit was very significantly improved when a second 
component was added. A two-temperature model yielded ${\mathrm k}T_{1} \simeq 
10.1 \pm 0.6$, ${\mathrm k}T_{2} \simeq 2.8 \pm 0.6$, and $Z_{\mathrm{Fe}} \sim 
0.23 \pm 0.03$, a value consistent with previously measured values. An 
equally good fit was obtained by a combination of primary thermal and 
secondary NT components, with ${\mathrm k}T \simeq 8.9 \pm 0.6$, and photon index 
$\Gamma \simeq 2.4 \pm 0.3$. Similar results were obtained when a 
joint analysis was performed of the {\sl{RXTE}} data and archival ASCA data. 
The deduced value of $\Gamma$ is consistent with the measured 
spectrum of extended radio emission in A2319. Identification of the 
power-law emission as Compton scattering of the radio-emitting electrons 
by the CMB resulted in $B_{\mathrm{rx}} \sim 0.1-0.3$ $\mu$G, and $\sim 4 \times 
10^{-14} (R/2\,{\mathrm{Mpc}})^{-3}$ erg\,cm$^{-3}$ for the mean energy density of 
the emitting electrons in the central region (radius $R$) of the cluster.

\ni{\bf{A~2256:}} Following a very short $\sim 30$ ks observation of 
A~2256 (in 1997) - which resulted in an upper limit on NT emission 
\citep{henriksen1999} - the cluster was observed (in 2001 and 2002) for a 
total of $\sim$343\,ks ({\sl PCA}) and $\sim 88$ ks ({\sl HEXTE}). 
The data analysis \citep{rephaeli2003} yielded evidence for two 
components in the spectrum. Based on statistical likelihood alone the 
secondary component can be either thermal or power-law. Joint analysis 
of the {\sl{RXTE}} and archival {\sl{ASCA}} data sets yielded ${\mathrm k}T_1 = 
7.9^{+0.5}_{-0.2}$ and ${\mathrm k}T_2 = 1.5^{+1.0}_{-0.4}$, when the second 
component is also thermal, and ${\mathrm k}T = 7.7^{+0.3}_{-0.4}$ and 
$\Gamma = 2.2^{+0.9}_{-0.3}$ (90~\% CL), if the second component 
is a power-law. Identifying the secondary emission as due to Compton 
scattering of the radio producing relativistic electrons yielded $B_{\mathrm{rx}} 
\simeq 0.2^{+1.0}_{-0.1}$ $\mu$G in the central 1$^{o}$ region 
of the cluster, a region which contains both the halo and relic sources.

\ni{\bf{A~2163:}} The moderately-distant ($z=0.203$) cluster was 
observed for $\sim$530 ks (during a 6 month period in 2004). Primary 
thermal emission in this cluster comes from very hot IC gas with ${\mathrm k}T 
\sim 15$ keV, but analysis of the observations \citep{rephaeli2006} 
indicated very clearly that this component does not by itself 
provide the best fitting model. A secondary emission component was 
needed, and while this could also be thermal at a temperature 
significantly lower than $15$ keV, the best fit to the full 
dataset was obtained with a power law secondary spectral component. 
The parameters of the NT emission imply a significant fractional flux 
amounting to $\sim 25$~\% of the integrated $3-50$ keV emission. NT 
emission is expected given the intense level of radio emission, most 
prominently from a large radio halo. \citet{rephaeli2006} assumed that 
the NT emission originates in Compton scattering of (the radio-emitting) 
relativistic electrons by the CMB, and estimated $B_{\mathrm{rx}} \sim 0.4 
\pm 0.2\, \mu$G to be an overall mean field strength in the large 
complex region of radio emission in the cluster.

\ni{\bf{A~3667:}} The cluster was observed (in 2001 and 2002) for 
$\sim 141$ ks; analysis of the RXTE observation and lower energy {\sl{ASCA}} 
data, yielded only marginal evidence for a secondary power-law emission 
component in the spectrum \citep{rephaeli2004}. This resulted in 
an upper limit of $2.6 \times 10^{-12}$ erg\,cm$^{-2}$\,s$^{-1}$ (at 90~\% 
CL) on NT emission in the $15-35$ keV band. When combined with the measured 
radio flux and spectral index of the dominant region of extended radio 
emission, this limit implies a lower limit of $\sim 0.4$\,$\mu$G on the 
mean, volume-averaged magnetic field in A~3667.

\ni{\bf{1ES 0657-55.8:}} At $z=0.296$ the `bullet' cluster is the most 
distant cluster searched for NT emission with {\sl{RXTE}}. The cluster was 
observed (in 2002 and 2003) for a total of $\sim 400$ ks. Joint 
analysis of the {\sl{RXTE}} observations together with archival {\sl{XMM-Newton}} 
observations clearly indicated the presence of a second spectral 
component \citep{petrosian2006}. While the nature of this second component 
cannot be determined from the spectral analysis alone, the authors argue 
that a power-law spectral shape is a more viable interpretation than an 
exponential. Since no radio data were available, the authors could only 
determine a mean field value of $\sim 1.2$\,$\mu$G assuming energy 
equipartition between the field and relativistic electrons.

\subsection{Search for NT emission with {\sl BeppoSAX}}

For observations of clusters the two relevant instruments were the 
MECS and PDS, the former with a spectral range of $1.3-10$ keV, $56^{\prime}$ FOV 
and $\sim 2^{\prime}$ resolution, and the latter with response in the nominal 
$15-300$ keV range and $1.3^{\circ}$ FOV. The capability to separate the 
respective contributions of low and high energy components requires 
simultaneous and self-consistent measurements. For this reason the lack 
of spectral overlap of the two instruments (and their somewhat different 
FOV) was unfortunate.

\subsubsection{Observations of Individual Clusters}

Results of searches for cluster hard X-ray emission are summarised 
in Table \ref{hxr_sax.tab}; a brief discussion of these results follows.

\ni{\bf{Coma:}} \citet{fuscofemiano1999} reported a 4.5$\sigma$ 
detection of excess emission above the primary thermal emission in $\sim 
91$ ks PDS observations carried out in 1997, but a later re-analysis 
resulted in a reduced estimate of the significance to $3.4\sigma$ 
\citep{fuscofemiano2004}. The cluster was observed again 
for $\sim 300$ ks (in 2000), and an analysis of the combined datasets 
- with a thermal component whose temperature was {\sl assumed} to be 
$kT =8.1$ keV - yielded a $4.8\sigma$ detection of excess emission 
in the $20-80$ keV band \citep{fuscofemiano2004}. Assuming a power-law 
form for this excess emission with index $\Gamma \sim 2.0$, the 20-80 
keV flux of $1.5\pm0.5 \times 10^{-11}$ erg s$^{-1}$ cm$^{-2}$ was 
derived (a value which is consistent with that derived by \citet{nevalainen2004}. Interpreting this to be of Compton origin, and using the radio 
flux, resulted in $B_{\mathrm{rx}}\sim 0.2$ $\mu$G.

\begin{figure}
\begin{center}
\includegraphics[height=0.8\textwidth, angle=270]{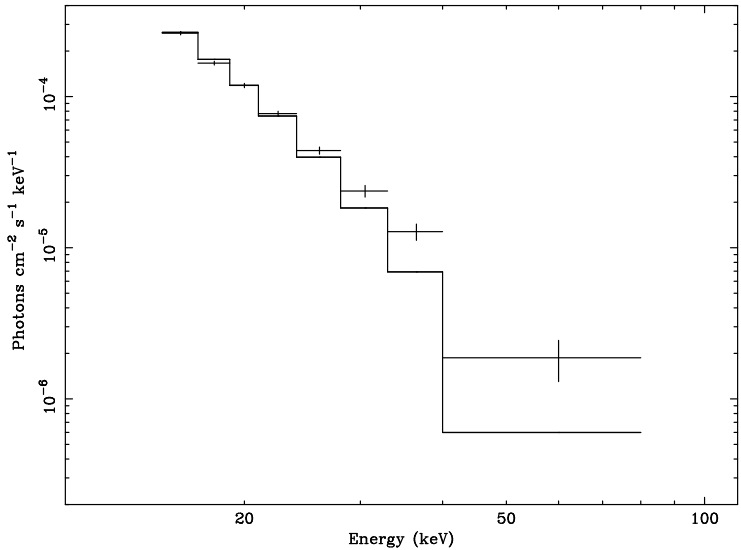}
\caption{{\sl{BeppoSAX}} spectrum of the Coma cluster obtained using the XAS 
software package (from \protect\citealt{fuscofemiano2007}). Data are from PDS 
measurements (above 15 keV, with 68~\% error bars); the line is a 
best-fit thermal model with $kT=8.11$ keV obtained previously from 
{\sl Ginga} measurements \protect\citep{david1993}.}
\label{fig:fig2}
\end{center}
\end{figure}

\citet{rossetti2004} questioned the reliability of the \citet{fuscofemiano1999,fuscofemiano2004} analyses. Having carried out their own analysis of 69 
blank fields observed with PDS, they found that the two offset positions 
used for the local background determination yielded systematically 
different results for the background-subtracted flux. Since \citet{fuscofemiano1999,fuscofemiano2004} used only one offset pointing (due to contamination 
in the other offset field), without any adjustment for the systematic 
difference, \citet{rossetti2004} claimed that the results of 
Fusco-Femiano et al. are incorrect. Based on the use of a different 
software package (SAXDAS) for background determination, \citet{rossetti2004} determined that the significance of detection of excess 
emission in the original observation was only at the $2\sigma$ level, 
while their analysis of the second observation yielded no evidence for 
excess emission, in strong contrast with the $4.8\sigma$ detection level 
deduced by \citet{fuscofemiano2004}. 

The analysis was repeated (see Fig.~\ref{fig:fig2}) by \citet{fuscofemiano2007}; they concluded 
that the discrepant results are due to the presence of two variable 
sources in one of the background fields, which is why they ignored that 
field in their analysis. The different background fields and the use of 
the XAS software (rather than SAXDAS) were suggested to be the reasons 
for the high significance ($4.8\sigma$) detection reported by 
\citet{fuscofemiano2004}. In a recent reply to \citet{fuscofemiano2007}, \citet{rossetti2007} defended their 
re-analysis and the lower detection significance of the hard excess, 
re-emphasising that the significance is highly sensitive to both the 
choice of the offset pointings for local background estimation, and also 
to the exact value of the temperature adopted in the analysis. 
\citet{landi2005} analysed a sample of 868 blank fields and found no 
difference in flux values when using either of the two offset 
pointings. However, \citet{rossetti2007} could not investigate 
the reasons for the discrepancy between their blank sky measurements 
and those of \citet{landi2005} since the details of the latter work 
are not yet publicly available. 

Flux confusion due to unidentified (and therefore unremoved) background 
AGN is particularly troublesome due to the relatively large FOV 
($1.3^{\circ}$) of the PDS. Although some of the integrated emission 
of distant AGN in the cluster field is removed in the rocking mode,  
there still is a residual signal due to AGN background fluctuations, and 
possibly also the presence of AGN in the cluster. This introduces a level 
of systematic uncertainty that was not specifically accounted for in most 
published PDS cluster analyses. Analysing 164 PDS blank-sky observations, 
\citet{nevalainen2004} found a systematic difference in the 
background-subtracted fluxes when using only one of the two pointings 
available for the local background estimate in the standard rocking mode. 
The authors speculate that this might be due to either radiation entering 
the collimators from the side, screening of the instruments by the 
satellite, or the fact that the detector is looking at more/less 
radioactive parts of the satellite. These results are qualitatively 
similar to those of \citet{rossetti2004}, but are more pronounced 
by a factor of $\sim 2$. The difference in flux values when using 
either of the two offset pointings seems to be the reason for the 
different results obtained for the significance of the detection of NT 
emission in Coma. The issue remains unsettled and needs to be further 
explored.

Moreover, as we have pointed out, precise separation between the primary 
and secondary spectral components depends very much on the ability to do 
a simultaneous fit to the parameters of both thermal and possibly NT 
emission. The search for a NT component with {\sl{BeppoSAX}} suffered from lack of 
spectral overlap between the MECS and PDS instruments, and consequently 
the need to {\sl adopt} a previously determined value of ${\mathrm k}T$. Due to 
inherent differences between the {\sl HEXTE} and {\sl PDS} instruments, 
only a rough comparison between the respective results can be made. In 
this spirit it is apparent that the two estimates for the NT ($20-80$ keV) 
flux \citep{rephaeli2002,fuscofemiano2004} are in agreement 
(see Tables 1 \& 2).

\ni{\bf{A~2256:}} \citet{fuscofemiano2005} analysed two separate 
PDS observations of the cluster for a total of $\sim 430$ ks; 
they claimed a $4.8 \sigma$ detection of excess emission. 
Fitting this excess to a power-law yielded 
$\Gamma = 1.5^{+0.3}_{-1.2}$ and a $20-80$ keV flux of 
$8.9\times 10^{-12}$ erg cm$^{-2}$  s$^{-1}$, which is 
essentially the same value deduced by \citet{nevalainen2004}. Use 
of the measured radio flux yielded $B_{\mathrm{rx}} \sim 0.05$ $\mu$G, 
assuming that the NT emission they deduced originates in the radio 
relic located in the NW side of the cluster. We note that the flux 
deduced by \citet{rephaeli2003} is about a factor of $\sim 2$ 
lower than that of \citet{fuscofemiano2005}, but given the large 
uncertainty in $\Gamma$, for which the former authors quote a 
relatively low value, this difference is not very significant.

\ni{\bf{A~2199:}} Analysis of MECS $8-10$ keV measurements led \citet{kaastra1999} to claim detection of a significant excess emission (over 
that of the best-fit thermal model). Fit of a power-law to the data 
yielded an index of $1.81\pm 0.25$. They concluded that this emission 
originates in an outer region (beyond a 300 kpc central radial region), 
and that its relative strength (with respect to thermal emission) 
increases such that it dominates the $8-10$ keV emission at $\sim 1$ Mpc 
from the centre.

\begin{table}
\caption{NT emission parameters from {\sl{BeppoSAX}} measurements
and (deduced mean radio-and-X-determined field) $B_{\mathrm{rx}}$}
\label{hxr_sax.tab}
\begin{tabular}{lllll}
\hline
Cluster & $20-80$ keV flux                        & $\Gamma$         & $B_{\mathrm{rx}}$  & Reference\\
        & ($10^{-12}$ erg s$^{-1}$ cm$^{-2}$) &                   & ($\mu$G)  &\\
\hline
Coma$^{\bf a}$ & $15 \pm 5$          &   ?           & 0.2        & \protect\citet{fuscofemiano2004}\\
A~2256           & $8.9^{+4.0}_{-3.6}$        & 1.5$^{+0.3}_{-1.2}$      & 0.05 & \protect\citet{fuscofemiano2000}\\
A~2199           & $9.8 \pm 4.0$                     & 1.8$\pm$0.4   & ?           & \protect\citet{kaastra1999}\\
A~2319           & $\le 23$            & ?             & $\ge$0.04  & \protect\citet{molendi1999}\\
A~3667$^{\bf b}$ & $\le$6.4            & 2.1            & $\ge$0.4   & \protect\citet{fuscofemiano2001}\\
A~754$^{\bf c}$  & $\sim 2$           & ?              & 0.1         & \protect\citet{fuscofemiano2003}\\
Centaurus$^{\bf d}$  & $10.2 \pm 4.6$             & 1.5$^{+2.3}_{-1.3}$& & \protect\citet{molendi2002}\\
\hline
\multicolumn{5}{l}{All quoted errors are at the 90~\% confidence level.} \\
\multicolumn{5}{l}{$^{\bf a}$ According to Rosetti \& Molendi (2007) proper accounting 
for uncertainties}\\
\multicolumn{5}{l}{ makes the detection insignificant.}\\
\multicolumn{5}{l}{$^{\bf b}$ The emission from AGN FRL339 was not subtracted out.}\\
\multicolumn{5}{l}{$^{\bf c}$ The measured flux is possibly from the radio galaxy 26W20.}\\
\multicolumn{5}{l}{$^{\bf d}$ The measured flux is possibly due to an AGN.}\\
\end{tabular}
\end{table}

\ni{\bf{A~3667:}} A marginal detection (at a $2.6\sigma$ significance) 
of excess $20-35$ keV emission was reported by \citet{fuscofemiano2001}. 
A power-law fit resulted in an index of $\sim 2.1$, from which an upper 
limit on NT $20-80$ keV flux of $6.4 \times 10^{-12}$ 
erg\,s$^{-1}$\,cm$^{-2}$, and a lower limit $B_{\mathrm{rx}} \simeq 0.4$ $\mu$G 
were deduced. However, the authors did not consider the impact of the 
presence of the Seyfert 1 galaxy FRL 339 in the PDS FOV; emission from 
this galaxy is at a level comparable to that attributed to the NT 
component \citep{nevalainen2004}. This would seem indeed the case, 
given the fact that analysis of {\sl{RXTE}} measurements yielded only an upper 
limit on NT emission \citep{rephaeli2004}.

\ni{\bf{A~754:}} \citet{fuscofemiano2003} reported the detection of 
excess emission above 45 keV at a $3.2\sigma$ significance. They deduced 
a $10-40$ keV flux of $2\times 10^{-12}$ erg\,s$^{-1}$\,cm$^{-2}$ and 
$B_{\mathrm{rx}} \simeq 0.1$ $\mu$G, but noted that the presumed NT emission could 
possibly be from the radio galaxy 26W20.

\ni{\bf{Centaurus:}} \citet{molendi2002} detected a hard X-ray excess 
at the $3.6\sigma$ level. The best-fit power-law model yielded an index 
of $\Gamma = 1.5^{+1.4}_{-0.8}$, and a $20-200$ keV flux of $2.2 \times 
10^{-11}$ erg\,s$^{-1}$\,cm$^{-2}$, but they concluded that the emission 
may originate in a serendipitous AGN.

\ni{\bf{A~2319:}} No statistically significant power-law emission was 
detected in the analysis of a very short $\sim 20$ ks (PDS) observation, 
resulting in an upper limit of $2 \times 10^{-11}$ erg\,s$^{-1}$\,
cm$^{-2}$, and a lower limit $B_{\mathrm{rx}} \sim 0.04$ $\mu$G \citep{molendi1999}. As discussed above, the much longer ($\sim 160$ ks) {\sl{RXTE}} 
observation led to a significant NT flux.

\subsubsection{Statistical results from a cluster sample} 

An attempt to obtain some insight from co-added PDS data on a sample of 27 
clusters was made by \citet{nevalainen2004}. A brief review of their 
results follows.

The problematic aspects of a statistical study of PDS data should first 
be summarised. As we have already noted, flux confusion due to 
unidentified AGN is of particular concern due to the relatively large PDS 
FOV. \citet{nevalainen2004} used optical catalogues to identify Seyfert 
1 galaxies, and BeppoSAX MECS or ROSAT PSPC data at their locations to 
estimate the AGN contribution in the PDS spectra. Clusters for which the 
estimated Seyfert 1 contribution was more than 15~\% of the total signal 
were removed from the initial sample. Population synthesis modelling of 
the cosmic X-ray background (CXB) indicates that 80~\% of the AGN need to 
be obscured to produce the CXB spectrum \citep{gilli1999}, which is 
harder than the spectrum of unobscured AGN. Indeed, recent deep X-ray 
observations of blank fields (e.g., \citealt{hasinger2001}) have discovered 
a population of absorbed point sources that outnumbers the Seyfert 1 
galaxies by a factor of $\sim$4. These obscured AGN are seen through an 
absorbing torus (with $N_{\mathrm H} = 10^{22} - 10^{25}$ atoms\,cm$^{-2}$; \citealt{risaliti1999}) which hides them in the soft X-ray band. Thus, a robust 
estimate for their contribution was not available. However, the spectral 
and spatial distribution of the NT emission (see below) argues against 
significant contamination due to obscured AGN in the final \citet{nevalainen2004} sample.

\begin{figure}    
\begin{center}
\includegraphics[width=9.cm,clip=]{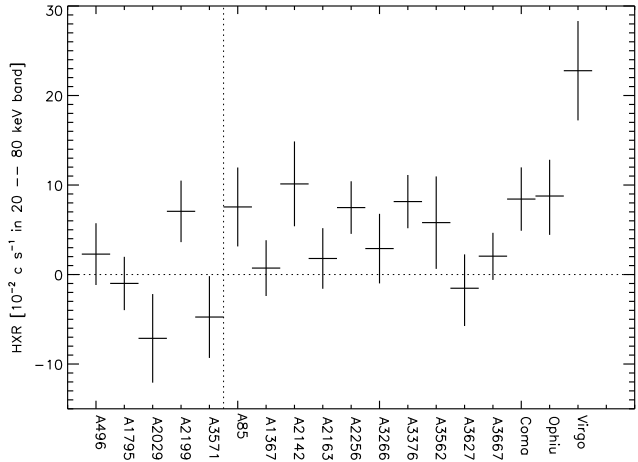}
\caption{NT signal and 1$\sigma$ uncertainties in the PDS $20-80$ keV band 
after subtraction of the contributions from the background, thermal 
emission, and AGN in the field, and accounting for uncertainties in these 
subtractions \protect\citep{nevalainen2004}. The dotted vertical line 
separates the relaxed clusters (left) from the rest (right).}
\label{fig:fig3}
\end{center}  
\end{figure}

The thermal emission of clusters in the sample was modelled by \citet{nevalainen2004} based on results of published analyses of the emission 
measured by the BeppoSAX MECS and XMM-Newton EPIC instruments. Since the PDS has 
no spatial resolution, the flux observed in the central region was 
extrapolated to the full PDS FOV using the appropriate single (or double) 
$\beta$ models, after account was made for vignetting. They extrapolated 
the thermal emission model to the $20-80$ keV band and compared the 
prediction with the PDS data, thus estimating the NT emission. The 
comparison showed that in $\sim$50~\% of the clusters the NT component 
was marginally detected at the 2$\sigma$ level. Most of the significant 
detections were found in clusters which show signs of recent merger 
activity. Specifically, when clusters were divided into `relaxed' or 
`merger' groups, it was determined that the mean $20-80$ keV NT emission 
in the latter group ($4.8 \times 10^{-2}$ 
counts\,s$^{-1}$) was $\sim 10$ times higher than that of the `relaxed' 
clusters (see Fig.~\ref{fig:fig3}). The level of 
systematic uncertainties due to background fluctuations is such that 
only the emission from clusters in the `merger' group was found to 
be marginally significant (at $\sim 2\sigma$). Assuming a power-law 
shape for this excess with (photon) index of 2.0, it was deduced that 
the $20-80$ keV luminosity per cluster is $4 \times 10^{43} 
h_{70}^{-2}$ erg\,s$^{-1}$ \citep{nevalainen2004}.

\begin{figure}[!t]   
\begin{center}
\includegraphics[width=0.7\textwidth,clip=]{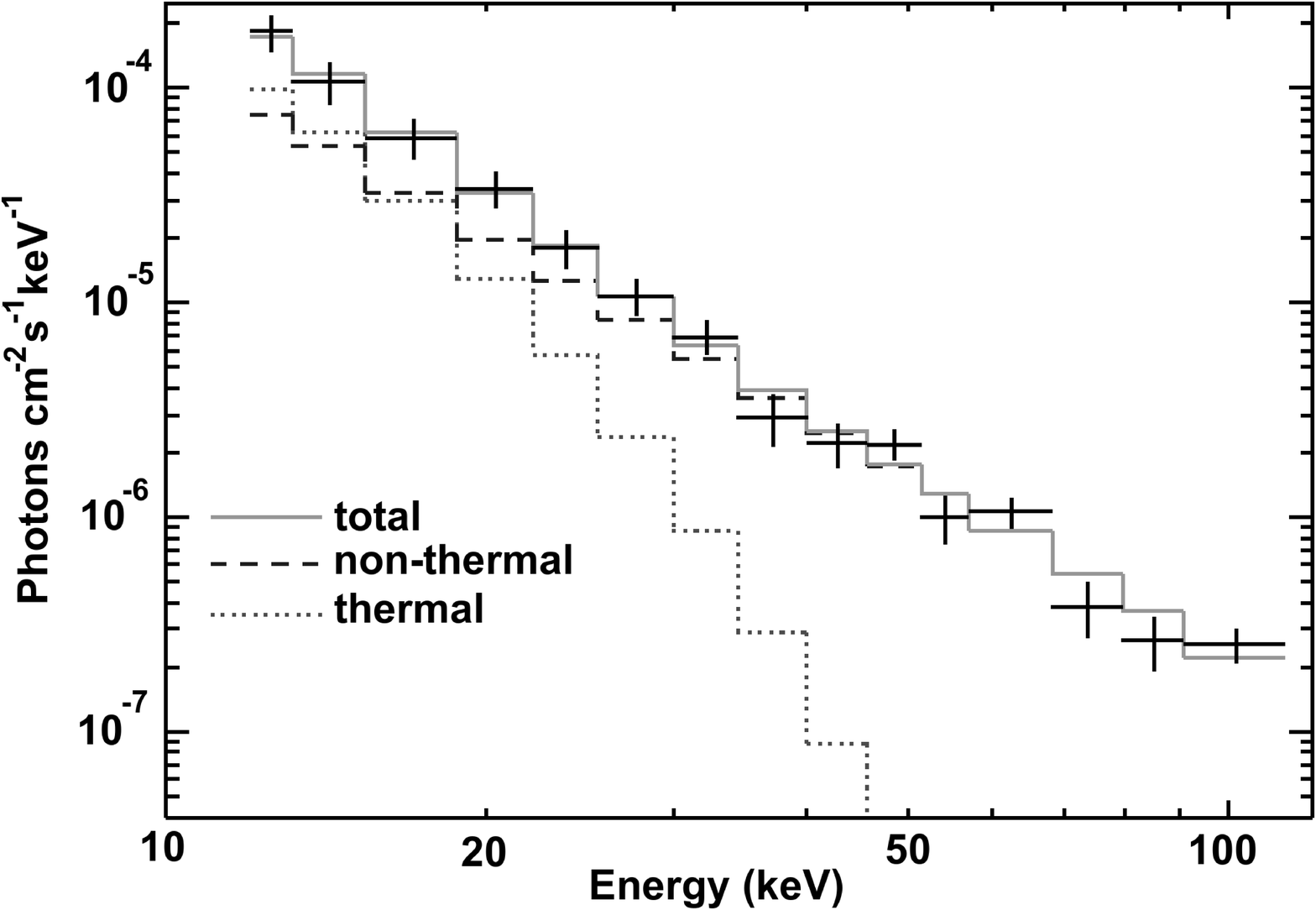}
\caption{The combined PDS spectrum of all the clusters not significantly 
affected by AGN \citep{nevalainen2004}. Lines show the unfolded 
model components while crosses show the data and 1$\sigma$ errors. The 
dotted line describes the thermal component, and the solid line shows the 
combined thermal and NT components. The dashed line indicates the 
best-fit power-law of $\Gamma = 2.8$.}
\label{fig:fig4}
\end{center}  
\end{figure}

The residual emission of the individual clusters was too weak for 
meaningful spectral analysis; therefore, \citet{nevalainen2004} 
co-added the data for the clusters, which amounted to a total of 560 ks. 
In order to obtain an estimate for the thermal contribution, the authors 
fitted data in the $12-20$ keV band obtaining a best-fit temperature of 
$\sim 8$ keV, consistent with the sample median. Extrapolating this 
model to the $20-80$ keV band showed that there was indeed excess emission 
(see Fig.~\ref{fig:fig4}). A fit of the full $12-115$ keV 
data to a thermal model resulted in a statistically unacceptable fit 
with an unrealistically high temperature of $\sim 26$ keV. 

A fit of the residual emission with a power-law model yielded a best-fit 
index $\Gamma = 2.8^{+0.3}_{-0.4}$ at 90~\% CL. Since a typical 
AGN index is $\sim 1.8$, appreciable AGN emission was ruled out at 
the 98~\% CL. The high level of NT emission in the $12-20$ keV band 
($\sim50$~\% of the total) is problematic, because in the central 
regions of the clusters in which NT emission was deduced, the MECS 
data typically allow the fractional contribution of such emission 
to be only a few percent. This seems to suggest that NT emission is 
extended with only a small fraction originating in the central region.

The indication from these statistical results for a connection between the 
cluster merger state and NT emission may be seen to support the scenario 
whereby electrons are accelerated by merger shocks. If so, the deduced 
power-law indices correspond to differential relativistic electron spectra with 
indices in the range $\mu = 3.8-5.0$. The implied radio synchrotron 
(energy) spectral indices $\sim 1.4-2$, are consistent with the 
observed range. 

\subsection{Search for NT emission with {\sl ASCA}}

{\sl ASCA} had a fairly good sensitivity and low background up to about 10 keV,
clearly too low for observing cluster NT emission. But there  could also be
detectable NT emission in groups of galaxies, in which  galaxy interactions
occur at relatively high encounter rates, and  given that gas temperatures in
groups are typically about 2 keV or  less, an attempt was made to look for NT
emission with {\sl ASCA}.  The first reported detection of excess emission, at
energies above  5 keV, was made for the group HCG~62 \citep{fukazawa2001}. The 
luminosity of the excess component was $\sim 4 \times 10^{41}$  erg s$^{-1}$,
which was estimated to be some 20 times higher than  the contribution of
discrete X-ray sources. The hard excess was  spatially extended to about $10'$
from the centre. Since the spectrum  was too hard to be interpreted as thermal
emission from the intra-group  gas, a NT origin was thought to be more likely. A
recent analysis of  {\sl Suzaku} XIS and HXD (see below) measurements of HCG~62
resulted only  in an upper limit on NT emission \citep{tokoi2007}, but at a
level which  does not exclude the ASCA result.

\begin{figure}
\begin{center}
\includegraphics[height=0.7\textwidth,angle=-90]{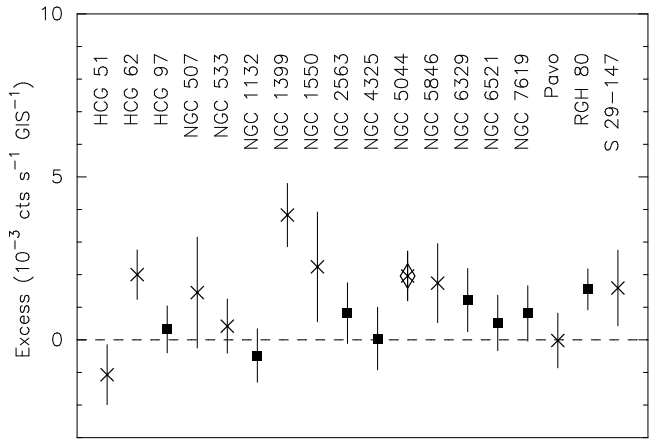}
\caption{{\sl{ASCA}} measurement of excess emission in a sample of 18 
groups of galaxies. The count rate in the $4-8$ keV band is shown by
the residual over the thermal model determined in the energy range
below 2.5 keV. Squares denote groups for which 
the soft-band spectrum was fitted with single temperature model, while 
in the systems shown by stars two-temperature models were required. 
Error bars include statistical and systematic $1\sigma$ uncertainties. 
A thermal fit to the NGC~5044 spectrum, indicated with a diamond, was 
considered to be unacceptable. Adapted from \protect\citet{nakazawa2007}.}
\label{fig:fig5}
\end{center}
\end{figure}

\citet{nakazawa2007} carried out a search for NT emission from 18 
groups of galaxies observed with {\sl{ASCA}}, including HCG~62. They 
fitted the spectra below 2.5 keV with 2 temperature thermal (MEKAL) 
models and compared the data in the $4-8$ keV band with the extrapolated 
thermal spectra. Excess fluxes are shown in Fig.~\ref{fig:fig5} in terms of statistical significance. 
HCG 62 and RGH 80 show excess emission at $> 2\sigma$ CL, with 
excess emission thought to be likely also in NGC~1399. The residual 
spectra could be fitted either with a thermal model with 
${\mathrm k}T \geq 3$ 
keV or a power-law model with photon index fixed at 2. The $2-10$ 
keV luminosity of the excess component is $10-30$~\% of the thermal 
emission and $4-100$ times stronger than the contribution from discrete 
X-ray sources. They concluded that both thermal and NT origins 
are acceptable from a statistical point of view.

\subsection{{Search for NT emission with \sl Suzaku}}
\label{Suzaku} 

Suzaku, the 5th Japanese X-ray satellite launched in July 2005, carries 
hard X-ray detectors (HXD) along with the X-ray CCD instrument (XIS). 
The combined energy range is from 0.3 keV up to about 600 keV, and 
a wide-band coverage of X-ray and $\gamma$-ray sources is possible. 
The HXD system consists of 16 units of well-type phoswich detectors,
surrounded by anti-coincidence shield scintillators \citep{takahashi2007,kokubun2007}. The well-type detector consists of a long
active collimator made of BGO (a Bismuth-Germanium Oxyde), and the bottom of the well is equipped
with Si PIN detectors and a GSO (a Gadolinium-Silicon Oxyde) scintillator. The PIN is 2 mm thick
and sensitive from 8 keV to 50 keV; the GSO covers the $50-600$ keV 
band. Typical effective areas are 160 cm$^2$ at 10 keV (PIN), and
330 cm$^2$ at 100 keV (GSO). The FOV is $34' \times 34'$ FWHM below 100 
keV, limited by a phosphor bronze fine collimator installed in the well. 
At higher energies, the FOV becomes wider, up to $4.5^\circ \times 
4.5^\circ$. The PIN detectors are cooled to $-20^\circ$C by a thermal 
radiator connected with heat pipes.

The background level is very low for the PIN detector, since it has no
accumulated effect of radio-activation in orbit. The measured
background level in orbit is indeed lower than the level of the PDS 
(on BeppoSAX) by a factor of about 3, even though Suzaku's inclined 
orbit makes it more susceptible to the intense X-rays during passage 
through the radiation belt. This and the HXD narrow FOV make it the 
most sensitive instrument in the $10-60$ keV band, among all previously flown hard X-ray 
detectors.

About 30 clusters have already been observed (as of April 2007) with 
Suzaku, including several for which the primary purpose is a search 
for NT emission. Meaningful constraints on NT emission were obtained 
for rather low temperature clusters. Most extensively observed is A~3376; 
these observations are described in some detail.

\ni{\bf{A~3376:}} This nearby ($z=0.046$) merging cluster has an IC gas 
temperature of $\sim 4$ keV. The relatively low temperature motivated 
the selection of the cluster as the prime target for observations with 
the PIN detectors. The cluster is noted by two large radio relics, 
probably expanding over a Mpc scale in the east and west boundaries of 
the thermal emission. BeppoSAX observations resulted in a $2.7\sigma$ 
detection of a NT component. The Suzaku observations were carried out 
during two separate pointings in October and November 2005. The first 
$\sim90$ ks observation covered the central region which includes 
the east relic, and the second $\sim 103$ ks was centred on the 
west relic region.

The background properties have been investigated by \citet{kawano2007} 
and applied to the A~3376 data. Background data, i.e.\ data taken when 
the satellite was pointing to the dark Earth, were sorted by position of 
the satellite in Earth coordinates. This basically performs sorting with 
the cosmic-ray cut-off rigidity which is known to correlate well with the 
non X-ray background. Since passage through the radiation belt leaves 
enhanced emission afterward, the data in each Earth position were 
further divided into northward and southward satellite movements. The 
long-term trend of the background variation was also included in the 
estimation. Comparison with the predicted background rate and the 
observed dark Earth data indicates that this method gives 3.5~\% 
($1\sigma$) error.

\begin{figure}
\begin{center}
\includegraphics[width=12cm,angle=0]{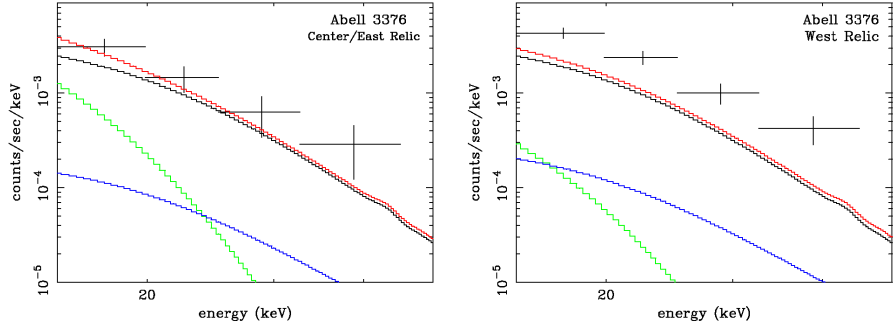}
\caption{Background subtracted spectra of A~3376 with the {\sl{Suzaku}} HXD PIN 
detector \protect\citep{kawano2007}. Left panel is for the region including 
the centre and the east relic; the right panel is for the west relic. 
Upward along the left-hand ordinate, 
the four solid curves show 
estimated flux due to point-sources, 
cluster thermal emission, 
cosmic X-ray background contribution, 
and their sum, respectively.} 
\label{fig:fig6}
\end{center} 
\end{figure}

This method was applied to the observed data in the energy range $15-50$ 
keV. Thermal cluster emission, estimated from the XIS measurements, 
CXB and point source contributions need to be subtracted. 
As shown in Fig.~\ref{fig:fig6}, the HXD PIN spectrum for 
the west relic suggests some level of excess emission. 
Including fluctuations of all these components, the flux of the west 
relic in the $15-50$ keV band is $(6.3\pm 1.8\pm 6.2)\times 10^{-12}$ 
erg\,cm$^{-2}$\,s$^{-1}$, where the errors are statistical and systematic, 
respectively. Most of the systematic error, about 85~\%, is non-CXB; 
thus, further improvement in the background estimation is necessary 
in order to better determine the level of excess emission. The implied 
upper limit, $1.4 \times 10^{-11}$ erg\,cm$^{-2}$\,s$^{-1}$, is 
consistent with the limit obtained from the {\sl{RXTE}} measurements, and 
about 20~\% lower than the level reported from the {\sl{BeppoSAX}} observation.

The west-relic region was jointly observed with the XIS instrument,
and Kawano et al. also constrained the possible power-law flux based
on the spectral fit. The obtained upper limit in the $4-8$ keV region
is extrapolated to $15-50$ keV assuming a photon index of 1.8 with
no scaling with the FOV, and it corresponds to about 20~\% of
the hard X-ray upper limit. Of course, it is quite possible that the
true NT flux is less than $3\times 10^{-12}$ erg\,cm$^{-2}$\,s$^{-1}$ 
in the $15-50$ keV band; however, if the emission is spatially extended then the 
XIS constraint still allows a hard flux of $2.4\times 10^{-11}$ 
erg\,cm$^{-2}$\,s$^{-1}$ over the PIN field of view. Thus, an extended 
hard X-ray emission (larger than $0.5^{\circ}$) remains a possibility.
Using the 1.4 GHz measurement of the west relic, Kawano et al. 
deduced a lower limit $B_{\mathrm{rx}} >0.1\, \mu$G.

\ni{{\bf{A~1060}} and the {\bf{Centaurus cluster:}}} These clusters 
are both relaxed nearby systems with $kT\sim 3.5$ keV; A~1060 is non-cD 
cluster, while the Centaurus cluster is dominated by a cD galaxy. The 
clusters were well studied by \citet{kitaguchi2007}. The observed 
data for the A~1060 and Centaurus cluster by the PIN instrument 
were accumulated for 28 and 26 ks, respectively. The latter authors 
applied the standard non-CXB subtraction and found that the blank-sky 
data yielded a fluctuation of $3-5$~\% ($1\sigma$) in the energy range 
$10-50$ keV. When the non-CXB flux was subtracted from the data, 
the data for both clusters showed agreement with the CXB above 20 
keV; the A~1060 data even showed a deficit of the counts. In the energy 
range $10-20$ keV, there remained an excess flux over the CXB 
level for both clusters. The flux was 4 or 8 times stronger than the 
CXB level for A~1060 and Centaurus, respectively. The cluster thermal 
emission was measured with {\sl{ASCA}}, which had a wider FOV than the PIN, 
and the measured temperatures were 3.3 and 3.8 keV for the two clusters. 
These spectra with the intensity consistent with the {\sl{ASCA}} measurement 
were compared with the PIN residual component. They found that the 
thermal emission can explain the PIN data quite well for both clusters, 
so only upper limits were derived on the NT emission.

\subsection{{\sl INTEGRAL} observation of the Coma cluster} 

\begin{figure}    
\begin{center}
\includegraphics[width=10.0cm,clip=]{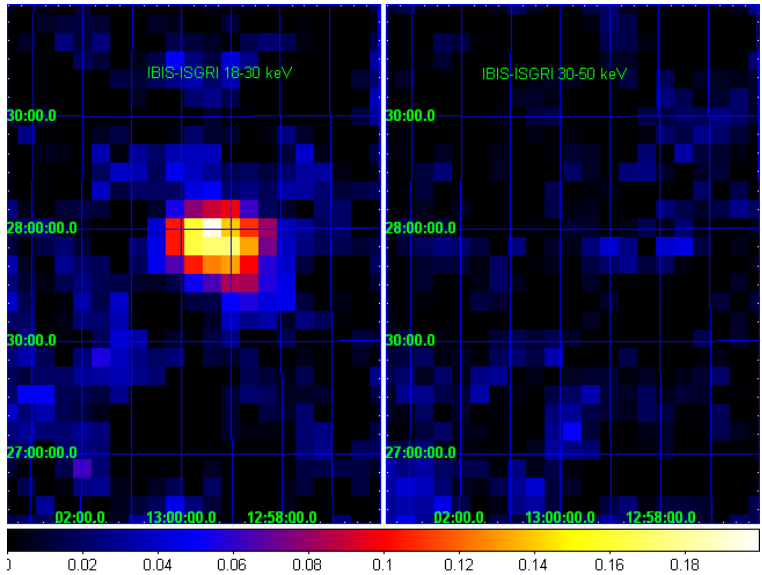}
\caption{Hard X-ray images of the Coma cluster from a 940 ks 
{\sl IBIS-ISGRI} observation. The left and right panels show the 
{\sl ISGRI} mosaics in the $18-30$ keV, and $30-50$ keV bands, respectively.}
\label{fig:fig7}
\end{center}  
\end{figure}

A large fully coded field of view ( $8^{\circ}\times 8^{\circ}$), and 
good imaging capabilities with a PSF of $12^{\prime}$ FWHM, make it possible to 
construct hard X-ray cluster images below 300 keV with the IBIS/ISGRI 
coded mask instrument \citep{ubertini2003,lebrun2003} 
aboard the {\sl International Gamma-Ray Astrophysics Laboratory} 
({\sl INTEGRAL}, \citealt{winkler2003}). \citet{renaud2006}  reported the 
analysis of a $\sim 500$ ks observation of the Coma cluster with IBIS/ISGRI. 
The cluster was detected at a significance $\sim 10\sigma$ CL in the 
$18-30$ keV energy band. The ISGRI image shows extended emission structure 
globally similar to that is seen by XMM-Newton below 10 keV (e.g.
\citealt{arnaud2001,neumann2003}). In Fig.~\ref{fig:fig7} we present 
two images of the Coma field in the $18-30$ keV (left panel) and $30-50$ keV 
(right panel) energy bands processed with the OSA-5 software, following a  
procedure similar to that described in \citet{renaud2006}, but for a 
longer (accumulated) ISGRI exposure of $\sim 940$ ks (`good time'). The 
detected flux in the $18-30$ keV band is consistent with the extrapolation of 
the flux measured by {\sl XMM-Newton} below 10 keV assuming a model of 
a thin thermal plasma model with ${\mathrm k}T=8$ keV (see \citealt{renaud2006,eckert2007}); no statistically significant flux is seen in 
the higher energy band. From the observed images it appears that in 
order to detect or meaningfully constrain with {\sl INTEGRAL} the 
presence of a NT X-ray component reported in {\sl RXTE} and {\sl 
BeppoSAX} observations of Coma, an integration time longer by a factor 
of $\sim 2-3$ would be required.

\subsection{Limits on cluster $\gamma$-ray emission}

The relatively steep power-law shape of cluster NT emission (which is 
most precisely measured in the radio) directly implies that the detection 
of $\gamma$-ray emission from even nearby clusters will be very 
challenging. To estimate the predicted level of emission, consider 
the Coma cluster as an example; assuming no change in the relativistic 
electron power-law index, and the deduced  $20-80$ keV flux level (in 
Table \ref{hxr_xte.tab}), the cumulative flux at $\epsilon \geq 
100$ MeV is $\sim 6\times 10^{-9}$ cm$^{-2}$\,s$^{-1}$. Although 
substantially uncertain, flux at this level is close to the projected  
{\sl GLAST} sensitivity threshold for an exposure time of 1 yr for the all sky survey. 
The flux will be well below this level if the electron spectrum steepens 
even modestly at energies above the range deduced from radio 
measurements. Note that $\gamma$-ray emission could also have hadronic 
origin (from $\pi^{0}$ decay, following proton-proton interactions), 
but only very rough limits can be placed on this emission (e.g., \citealt{dermer1988}). 

A statistical upper limit on $\gamma$-ray emission from clusters was 
obtained by \citet{reimer2003} from analysis of {\sl EGRET} 
measurements. The emission in fields centred at 58 clusters 
observed between 1991 and 2000 was analysed. A co-added mean flux 
level was determined after accounting for the diffuse background 
contribution. The resulting upper 2$\sigma$ limit on the mean cluster 
flux above (photon) energy of $\epsilon \geq 100$ MeV was found to be 
$6\times$10$^{-9}$ cm$^{-2}$\,s$^{-1}$. Interestingly, this level is 
at the level predicted for Coma based on direct extrapolation of 
the deduced $20-80$ keV flux. This {\sl EGRET} upper limit was used by 
\citet{bykov2000} and \citet{petrosian2001} to constrain models of hard 
X-ray emission components of leptonic origin; similarly \cite{blasi1999} used it to constrain the energetic proton energy
density in IC space.
  
\section{Discussion}

It has been known all along that definite detection of NT X-ray 
emission in clusters is a challenging task. This is due to several 
factors, major among which are its intrinsically weak level - swamped 
as it is by the intense primary thermal emission - and the difficult 
to achieve requisite high sensitivity and low detector background in 
the $10-100$ keV band. As has been emphasised, results of the search 
are not unequivocal, even for Coma and A~2256 for which detection 
(by {\sl{RXTE}} and {\sl{BeppoSAX}}) was claimed to be at moderately high level 
of statistical significance. This is mainly due to source confusion and 
a complete lack of spatial information. The search should continue (with 
{\sl{Suzaku}} and future satellites), spurred by its sound physical basis - the 
ubiquity of the CMB, and the observed radio synchrotron emission 
from relativistic electrons. 

The expected level of NT emission of Compton origin depends steeply 
on the mean magnetic field strength in the central cluster region. 
Clearly, if this is typically a fraction of a $\mu$G, as has been 
deduced in the analyses described in the previous section, then 
prospects for more definite detection of NT emission - perhaps already 
with deep {\sl{Suzaku}} measurements - are indeed good. However, significantly 
higher $B_{\mathrm{fr}}$ values - a few $\mu$G - are deduced from FR measurements; 
had these been meaningful estimates of the {\sl volume-averaged} 
field strength, then definite detection of NT emission would have been 
seriously questioned.

As we have noted already (in Sect.~2.2), the apparent discrepancy 
between deduced values of $B_{\mathrm{rx}}$ and $B_{\mathrm{fr}}$ has been investigated 
at some length. These two field measures are quite different; the 
former is essentially a volume average of the relativistic electron density 
and (roughly) the square of the field, the latter is an average along 
the line of sight of the product of the field and gas density. All 
these quantities vary considerably across the cluster; in addition, 
the field is very likely tangled, with a wide range of coherence 
scales which can only be roughly estimated. These make the 
determination of the field by both methods considerably uncertain. 
Thus, the unsatisfactory observational status (stemming mainly from 
lack of spatial information) and the intrinsic difference between 
$B_{\mathrm{rx}}$ and $B_{\mathrm{fr}}$, do not allow a simple comparison of these 
quantities. Even if the large observational and systematic uncertainties, 
the different spatial dependences of the fields, relativistic electron density, 
and thermal electron density, already imply that $B_{\mathrm{rx}}$ and $B_{\mathrm{fr}}$ 
will in general be quite different. This was specifically shown by 
\citet{goldshmidt1993} in the context of reasonable assumptions 
for the field morphology, and the known range of IC gas density profiles. 
They concluded that $B_{\mathrm{rx}}$ is indeed generally smaller than $B_{\mathrm{fr}}$. 
The implication is that prospects for detection of NT emission should 
not be based on the relatively high deduced values of $B_{\mathrm{fr}}$.

In conclusion, {\sl{RXTE}} and {\sl{BeppoSAX}} measurements have yielded appreciable 
evidence for power-law X-ray emission in four clusters. These results 
motivate further measurements and theoretical studies of NT phenomena 
on cluster and cosmological scales.

\begin{acknowledgements}
The authors thank  ISSI (Bern) for support of the team ``Non-virialized 
X-ray components in clusters of galaxies''.
\end{acknowledgements}

\end{document}